\documentclass[preprint]{aastex}
\usepackage{epsf}
\usepackage{emulateapj5}
\usepackage{onecolfloat}
\usepackage{amsmath}

\newcommand{\etal}{{et al.~}}
\newcommand{\be}{\begin{equation}}
\newcommand{\ee}{\end{equation}}
\newcommand{\Mo}{{\rm M_\odot}}
\newcommand{\kpc}{\,{\rm kpc}} 
\def\E{{\cal E}}
\begin{document}
\submitted{The Astrophysical Journal, accepted}
\vspace{1mm}
\slugcomment{{\em The Astrophysical Journal, accepted}} 

\twocolumn[
\lefthead{GENERATING EQUILIBRIUM DARK MATTER HALOS}
\righthead{KAZANTZIDIS, MAGORRIAN \& MOORE}

\title{GENERATING EQUILIBRIUM DARK MATTER HALOS: \\
INADEQUACIES OF THE LOCAL MAXWELLIAN APPROXIMATION}

\vspace{1mm}

\author{Stelios Kazantzidis}
\affil{Institute for Theoretical Physics, University of Z\"urich, CH-8057 
Z\"urich, Switzerland}
\email{stelios@physik.unizh.ch}
\author{John Magorrian}
\affil{Theoretical Physics, Department of Physics,
University of Oxford, 1 Keble Road, Oxford U.K, OX1 3NP}
\email{magog@thphys.ox.ac.uk}
\author{Ben Moore}
\affil{Institute for Theoretical Physics, University of Z\"urich, CH-8057 
Z\"urich, Switzerland}
\email{moore@physik.unizh.ch}

%
%

\begin{abstract}
We describe an algorithm for constructing $N$-body realizations 
of equilibrium spherical systems. 
A general form for the mass density $\rho(r)$ is used, making it possible to 
represent most of the popular density profiles found in the literature, including 
the cuspy density profiles found in high-resolution cosmological simulations.
We demonstrate explicitly that our models are in equilibrium.
In contrast, many existing $N$-body realizations 
of isolated systems have been constructed under the assumption that the
local velocity distribution is Maxwellian. 
We show that a Maxwellian halo with an initial $\rho(r)\propto r^{-1}$ central density 
cusp immediately develops a constant-density core.  
Moreover, after just one crossing time the orbital anisotropy 
has changed over the entire system, and the initially isotropic model 
becomes radially anisotropic.
These effects have important implications for many studies, 
including the survival of substructure in cold dark matter (CDM) models.
Comparing the evolution and mass-loss rate of isotropic Maxwellian
and self-consistent Navarro, Frenk, \& White (NFW) satellites
orbiting inside a static host CDM potential,
we find that the former are unrealistically susceptible to tidal
disruption. Thus, recent studies of the mass-loss rate and disruption timescales
of substructure in CDM models may be compromized by using the 
Maxwellian approximation. We also demonstrate that a radially anisotropic,
self-consistent NFW satellite loses mass at a rate several times higher
than that of its isotropic counterpart on the same external tidal field and orbit.
\end{abstract}

]

\section{INTRODUCTION}
\label{sec:introduction}

Despite the inexorable increase in the dynamic range of current
cosmological $N$-body simulations, they still cannot address many 
questions regarding the detailed physical processes taking place
inside galaxies. Examples include the survival and evolution of 
substructure within dark matter (DM) halos 
(Moore, Katz \& Lake 1996; Taffoni \etal 2003; 
Hayashi \etal 2003), the effects of tidal stripping and dynamical friction 
on the orbits of satellites (Vel\'azquez \& White 1995; van den Bosch \etal
1999; Colpi, Mayer \& Governato 1999; Mayer \etal 2002) 
and the consequent heating of galactic disks 
(Quinn \& Goodman 1986; Vel\'azquez \& White 1999; 
Taylor \& Babul 2001; Font \etal 2001), the susceptibility of
disks to bar instabilities (Mihos, McGaugh \& de Blok 1997),
and the effects of these bars on halo
central density cusps (Debattista \& Sellwood 2000; 
Athanassoula 2002; Valenzuela \& Klypin 2003).
For applications like these, it is more appropriate to take a well-understood,
isolated, equilibrium galaxy or halo model and use $N$-body simulations to
investigate how it responds to appropriately applied perturbations.

Constructing an $N$-body realization of an isolated, equilibrium model
galaxy is a quite difficult task to accomplish in a controllable way.
There are two steps in constructing such a model:
(1) find the phase-space distribution function
(DF) of the desired model, and then (2) use Monte Carlo
sampling of this DF to generate the $N$-body realization.
The first step constitutes the most difficult part. Simple, analytical DFs are known for
only a handful of models, such as Plummer spheres (Plummer 1911), various
lowered isothermal models (e.g., King 1966), lowered power-law
models (Evans 1993; Kuijken \& Dubinski 1994), and a few special cases
(e.g., Jaffe 1983; Hernquist 1990; Dehnen 1993).
However, none of these models can provide a plausible description of the 
density profile $\rho(r)$ of a cold dark matter (CDM) cosmological 
halo. In order to generate more realistic models one has to
find numerically a steady state DF that reproduces the desired
density and internal velocity anisotropy profiles, and this is not trivial.

An attractive way to circumvent this difficulty is to use
the {\it local Maxwellian approximation}.
Instead of finding the DF numerically, the self-consistent velocity 
distribution at a given point in space is approximated by a 
multivariate Gaussian whose mean velocity and velocity dispersion tensor 
are given by the solution of the Jeans equations at this point.
A clear description of this technique is given in Hernquist (1993).
The advantage of this scheme is that it is relatively easy to implement 
and can straightforwardly be applied to composite, flattened models 
of galaxies (e.g., Springel \& White 1999; Boily, Kroupa \& Pe\~{n}arrubia 2001).  

The main aim of the present paper is to highlight a dangerous
shortcoming of this approximation when it is used to generate initial
conditions (ICs) for high-resolution numerical simulations.
Most interesting galaxy models have local self-consistent velocity profiles
that become strongly non-Gaussian especially near the model's center.
If one uses the local Maxwellian approximation 
to construct an $N$-body realization of such a model, the center of 
the resulting $N$-body system will be far from equilibrium.
As we illustrate below, when such a model is evolved in isolation,
it rapidly relaxes to a steady state whose density and 
velocity profiles differ significantly from the initial, intended ones. 
Therefore, we argue that any result based on an uncritical application of this
approximation should be treated with care.
 
As a clear manifestation of the consequences of this, we consider
tidal stripping of satellite galaxies orbiting inside a static host potential
and demonstrate that satellites constructed
using the local Maxwellian approximation can undergo rapid artificial 
tidal disruption.  This has important implications for the evolution
of substructure in CDM halos. For example, Taffoni \etal (2003) and Hayashi \etal (2003) 
used simulations of individual Maxwellian satellites orbiting within
a deeper CDM potential to study the rate of mass loss due to tidal stripping. The rate of mass
loss was high, and in some cases the satellites were completely disrupted. The implications
of these studies are important for comparisons of the observed satellite dynamics
with the predictions of CDM models (Stoehr \etal 2002).

This paper is organized as follows. In \S~\ref{sec:initial_conditions},
we describe a straightforward procedure for generating ICs for a wide variety of
spherical models with both isotropic and anisotropic velocity dispersion
tensors without resorting to the local Maxwellian approximation. 
In \S~\ref{sec:realizations}, we show that $N$-body models generated
using this procedure are indeed in equilibrium, whereas those generated 
using the local Maxwellian approximation are not.
To illustrate one implication of this, in \S~\ref{sec:evolution} we consider the 
tidal stripping of CDM substructure halos orbiting inside a more massive 
static host potential. We construct model satellites using each prescription and
compare their mass-loss rate and survival times.
Finally, we summarize our main conclusions in \S~\ref{sec:summary}.

\section{APPROXIMATION-FREE INITIAL CONDITIONS 
FOR SPHERICAL MODELS}
\label{sec:initial_conditions}

In order to construct realizations of $N$-body models,
we must initialize both the velocities and the positions of the particles, 
thus fully determining the ICs of the model. 
In principle, the procedure decribed here can be applied 
to any density profile whose DF satisfies the minimum requirement 
for a model to be physical, that is, to be everywhere non-negative. 

\subsection{Density Profile Assumptions}

We consider models with density profiles that can be fitted by the
general two-parameter formula (Zhao 1996; Kravtsov \etal 1998),
\be
\rho(r)=\frac{\rho_{\rm s}} {(r/r_{\rm s})^\gamma [1+(r/r_{\rm s})^\alpha]^
{(\beta-\gamma)/\alpha}} \qquad\hbox{($r \leq r_{\rm vir}$)} \ ,
\label{general_density}
\ee 
\vspace{1mm}
where $\gamma$ controls the inner slope of the profile,
$\beta$ the outer slope, and $\alpha$ the sharpness of the transition 
between the inner and outer profile. 
The general form of equation (\ref{general_density})
includes as special cases many of the popular
density profiles that are used to fit the halos found in cosmological
simulations. In this case, the characteristic inner density 
$\rho_{\rm s}$ and scale radius $r_{\rm s}$, are sensitive to the epoch 
of halo formation and tightly
correlated with the halo virial parameters, via the concentration, $c$, and 
the virial overdensity $\Delta_{\rm vir}$.
For example, the Navarro, Frenk, \& White (1996, hereafter NFW) 
density profile has $ (\alpha, \beta, \gamma)=(1,3,1)$. The
steeper profiles found in higher resolution simulations 
(Moore \etal 1999a; Ghigna \etal 2000; Jing \& Suto 2000; Klypin \etal 2001)
correspond to $(\alpha, \beta, \gamma)=(1.5,3,1.5)$.
In addition, the so-called $\gamma$-models (Dehnen 1993; 
Tremaine \etal 1994), which have proved very useful in the study of 
elliptical galaxies and 
bulges, can be represented by choosing $(\alpha, \beta, \gamma)=(1,4,\gamma)$
and $\rho_{\rm s} = (3-{\gamma})M / 4 \pi r_{\rm s}^3$, where $M$ is the
model's total mass.

Density profiles with outer slopes $\beta > 3$ lead to
finite mass models, but for
$\beta\le3$ the cumulative mass profile diverges as $r\rightarrow\infty$.
Of course, these model profiles are not valid out to arbitrarily large distances, 
but simply provide fits up to about the virial radius $r_{\rm vir}$.
Our goal is to find equilibrium models that
fit the profile (eq. [\ref{general_density}]) out to this radius. A sharp
truncation to $\rho=0$ for $r>r_{\rm vir}$ would result in unphysical
models with $f<0$.  Instead, we use an exponential cutoff for 
$r > r_{\rm vir}$, following Springel \& White (1999).
This sets in at the virial radius and turns off the profile
on a scale $r_{\rm decay}$, which is a free parameter and controls the
sharpness of the transition:
\begin{eqnarray}
\nonumber
\rho(r)=\frac{\rho_{\rm s}} {c^\gamma (1+c^\alpha)^
{(\beta-\gamma)/\alpha}} \left(\frac{r}{r_{\rm vir}}\right)^{\epsilon}
\exp\left(-\frac{r-r_{\rm vir}}{r_{\rm decay}}\right) \\
&\hspace{-1.8cm} (r>r_{\rm vir}) \ ,
\label{exp_cutoff}
\end{eqnarray}
where $c\equiv r_{\rm vir}/r_{\rm s}$ is the concentration parameter.
In order to ensure a smooth transition between equations (\ref{general_density})
and (\ref{exp_cutoff}) at $r_{\rm vir}$, we require the logarithmic slope 
there to be continuous. This implies 
\be 
\epsilon=\frac{-\gamma - \beta c^\alpha}{1+c^\alpha} +
\frac{r_{\rm vir}}{r_{\rm decay}}.
\label{eps}
\ee 
Note that depending on the 
adopted model and the value of  $r_{\rm decay}$,  
this procedure results in some additional mass beyond the virial radius.  
For example, for a Milky Way--sized halo model 
$(M_{\rm vir} \sim 10^{12} \,h^{-1} \Mo)$ with
$c=12$ and a choice of $r_{\rm decay}=0.1\ r_{\rm vir}$, 
the total halo mass is $\sim10 \%$ larger than $M_{\rm vir}$.

\subsection{Distribution Function Assumptions}

According to the Jeans  theorem (Lynden-Bell 1962; Binney \& Tremaine 
1987, hereafter BT87), the most general DF of an equilibrium spherical system  
can depend on the phase-space coordinates $({\bf r}, {\bf v})$ only through the
isolating integrals of motion: the binding energy per unit mass, 
$\E$, and angular momentum vector per unit mass, ${\bf L}$. 
Obviously there are many DFs of the form $f(\E, {\bf L})$ that can produce any given 
density profile $\rho(r)$. In this paper
we restrict our attention to a special class of 
non-rotating models with DFs of the form $f=f(Q)$, where
\be
Q\equiv\E-\frac{L^{2}}{2r_{a}^{2}}
\label{Q_param}
\ee
(Osipkov 1979; Merritt 1985a,b), with the additional constraint that $f(Q)=0$ 
for $Q\le 0$. In these models, the radial velocity dispersion $\sigma_r(r)$ 
is related to the tangential dispersions $\sigma_\theta(r)=\sigma_\phi(r)$ through
\be 
\beta(r)\equiv 1-{\sigma_\theta^2\over\sigma_r^2}={r^2\over r^2+r_{\rm a}^2}.
\ee
Here $r_{\rm a}$ is the anisotropy radius, which controls  
the degree of global anisotropy in the velocity distribution.
Inside $r_{\rm a}$ the velocity distribution is nearly isotropic,
while outside $r_{\rm a}$ it becomes increasingly more radially anisotropic.
In the limit $r_{\rm a}\rightarrow\infty$ the models reduce to isotropic
models with $Q$ corresponding to the total binding energy $\E$.
The density profile of such a model is 
\be
\rho(r)\equiv\int f(\E,L)\,\rm{d}^{3}{\bf v}.
\label{zero_mom}
\ee
The inversion of the above integral equation gives the DF
(Eddington 1916; BT87),
\be
f(Q) =\frac{1}{\sqrt{8} \pi^2}\left[ \int_{0}^{Q}
\frac{{\rm d}^2 \rho_{Q}}{{\rm d} \psi^2} \frac{{\rm d}
\psi}{\sqrt{Q-\psi}} + \frac{1}{\sqrt{Q}} \left (\frac{{\rm d} \rho_{Q}}
{{\rm d}\psi}\right)_{\psi=0}\right] \ ,
\label{Q_df}
\ee   
where $\rho_{Q}(r)\equiv\rho(r) (1+r^{2}/ r_{\rm a}^{2})$, and 
$\psi(r)=-\Phi(r)$ is the relative gravitational potential.
The second term of the right-hand side in equation (\ref{Q_df})
vanishes for any sensible behaviour of $\psi(r)$ and $\rho(r)$ at 
large distances. 
Note that the ${\rm d}^2\rho_Q/{\rm d} \psi^2$ factor in the
integrand would be difficult to deal with
numerically, but it can be evaluated analytically using the
expressions (\ref{general_density}) and (\ref{exp_cutoff}) for 
$\rho(r)$, to give an expression in which the only derivatives
remaining are ${\rm d}\psi/{\rm d} r$ and ${\rm d}^2\psi/{\rm d} r^2$.
Both of these can be written in terms of the density profile $\rho(r)$ and 
its cumulative mass distribution $M(r)$. Thus, equation (\ref{Q_df}) reduces to a 
simple quadrature, with no numerical differentiation required.

\subsection{Initialization Procedure}

Given a choice of parameters specifying the density profile $\rho(r)$,
we first calculate the model's cumulative mass
distribution $M(r)$ and gravitational potential $\psi(r)$ on a grid
between a minimum and a maximum radius. The grid points 
are spaced uniformly in $\log r$. The minimum radius, $r_{\rm min}$,
is chosen such that we are sufficiently close to the model's center 
(e.g., $r_{\rm min}=10^{-6}\ r_{\rm s}$). The choice of the maximum radius, 
$r_{\rm max}$, depends on the functional form of the adopted density profile. 
In the case of the finite mass models (e.g., $\gamma$-models), 
the maximum radius is chosen to enclose at least 99\% of the model's total 
mass, while in the case of the truncated density profiles 
we follow our models to a maximum radius equal to $r_{\rm vir}$ plus 
several $r_{\rm decay}$.
We use linear interpolation over the grid
whenever values of $M(r)$, $\psi(r)$, or ${\rm d}\psi/{\rm d} r$ are needed.  
In principle, it is straightforward to include the effects of the softening 
used in the $N$-body code in the results, but we have not found it 
necessary to do so for any of the applications in this paper.

The integral in equation (\ref{Q_df}) for the DF 
has an integrand with an integrable singularity at one
or both of its limits, but this is dealt with by a suitable change of
variables. The accuracy of the numerical integration is checked
by comparing its results against the exact analytical expressions that
exist for models with $(\alpha,\beta,\gamma)=(1,4,1)$ (Hernquist 1990)
and $(1,4,2)$ (Jaffe 1983). The fractional error in the numerical
calculation of the DF over the range of energies that correspond to
apocentric distances of radial orbits between $10^{-6}\, r_{\rm s}$
and $10^{6}\, r_{\rm s}$, was better than $10^{-5}$, except near the
edges of the grid where it increased to $10^{-3}$.

In practice, instead of evaluating equation (\ref{Q_df}) everytime a value of
$f(Q)$ is required, we first generate a table containing $f(Q_i)$
for a range of $Q_i$ equispaced in $\log Q$. We can then use linear
interpolation in $\log f$ and $\log Q$ to obtain values of $f(Q)$ for
any $Q$. We test the accuracy of this interpolation scheme by
evaluting the density integral of equation (\ref{zero_mom}) and comparing
with the exact expression. 
We find that the recovered density agrees very well with the 
original density, with fractional error typically better than $10^{-5}$, except 
near the edges of the grid where it rises to $10^{-4}$.
For all the models we have tried for this paper, we have found that the DF 
is everywhere non-negative and is an increasing function of $Q$.
Thus, the local velocity distributions $f({\bf v})$ always peak at ${\bf v} =0$  
which is important for the choice of particle velocities.

Once the density $\rho(r)$ has been specified and the cumulative mass
distribution $M(r)$, gravitational potential $\psi(r)$, and DF $f(Q)$
have been calculated we can start to randomly sample particles from
the DF. The particle positions are initialized using the
cumulative mass distribution $M(r)$. Having the particle's position, we
then use the acceptance-rejection technique (Press \etal
1996; Kuijken \& Dubinski 1994) to find its velocity.

\section{EVOLUTION OF ISOLATED N-BODY REALIZATIONS}
\label{sec:realizations}

We have described two ways of generating realizations of $N$-body 
models: the local Maxwellian approximation (\S~\ref{sec:introduction})
and the procedure utilizing the exact phase-space DF (\S~\ref{sec:initial_conditions}).
Here we demonstrate that models constructed using the procedure of 
\S~\ref{sec:initial_conditions} are in equilibrium, 
but that the Maxwellian models evolve significantly. In total we set up seven 
simulations:

\begin{enumerate}

\item {\it Model A}.---An isotropic halo of $3 \times10^{6}$ particles following the
Hernquist density profile. The ICs are generated using the local
Maxwellian approximation. All models, unless specified, have been constructed with 
$r_{\rm min}=10^{-6}\ r_{\rm s}$ and $r_{\rm max}=10^{2}\ r_{\rm s}$.

\item {\it Models B and C}.---Both models follow the Hernquist density profile with
$10^{5}$ particles and are evolved for much longer than
model A.  Each is constructed using the local Maxwellian
approximation. Model B is isotropic while model C is anisotropic, 
with anisotropy radius $r_{\rm a} =(4/3)\ r_{\rm s}$.

\item {\it Models D and E}.---Versions of models B and C, respectively,
constructed using the initialization procedure of \S~\ref{sec:initial_conditions} instead of
the local Maxwellian approximation.

\item {\it Model F}.---An isotropic halo of $10^{6}$ particles with an NFW
density profile, generated using the initialization procedure described in
\S~\ref{sec:initial_conditions}. This model represents a dwarf galaxy with 
$M_{\rm vir} = 10^{10} \,h^{-1} \Mo$ and $c=15$ and has been constructed 
with $ r_{\rm min}=10^{-6}\ r_{\rm s}$ and $r_{\rm max}=r_{\rm vir} + 3\ r_{\rm decay}$.

\item  {\it Model G}.---A halo similar to model F, but following the Moore
density profile (Moore \etal 1999a) and having a concentration of $c=9.5$.

\end{enumerate}

\begin{figure*}
\centerline{\epsfxsize=3.98in \epsffile{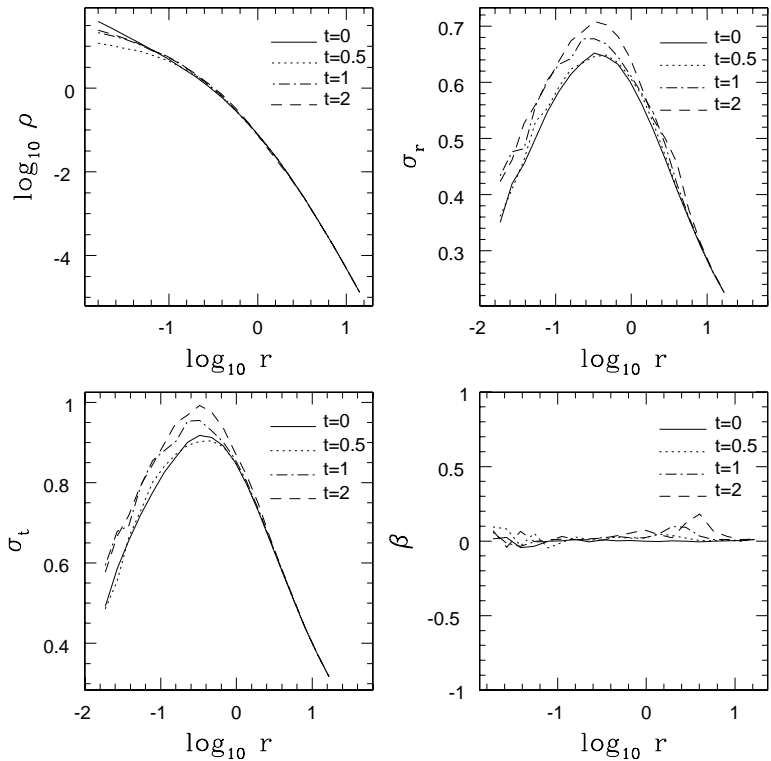}}
\caption {Initial evolution in the density profile ({\it top left}) and velocity structure (radial
dispersion [{\it top right}], tangential dispersion 
$\sigma_t^2=\sigma_\phi^2+\sigma_\theta^2$ [{\it bottom left}] and
anisotropy parameter [{\it bottom right}]) as a function of time for an
isotropic Hernquist model with $3\times10^6$ particles (model A). The
ICs were generated using the local Maxwellian approximation.  The
solid lines show the initial profiles. The dotted lines show the
profiles after $0.5$ crossing times at the scale radius. The profiles
after $1$ and $2$ crossing times at the scale radius correspond to the 
dot-dashed and dashed lines, respectively.
The velocity and density profiles evolve rapidly away from the initial
state, which is indicative of the fact that the model is not in equilibrium.
\label{fig1}}
\end{figure*}

We evolve our models using PKDGRAV, a multistepping, parallel 
$N$-body code written by J. Stadel \& T. Quinn (Stadel 2001).  
The code uses a spline softening length such that the force is completely 
Keplerian at twice the quoted softening lengths, with the equivalent 
Plummer softening being 0.67 times the spline softening.  
We used an adaptive, kick-drift-kick (KDK) leapfrog integrator, and the 
individual particle time steps $\Delta_{t}$ are chosen according to $\Delta_{t} \leq \eta
(\epsilon_{i}/\alpha_{i})^{1/2}$, where $\epsilon_{i}$ is the
gravitational softening length of each particle, $\alpha_{i}$ is the
value of the local accelaration, and $\eta$ is a parameter that
specifies the size of the individual time steps and consequently the
time accuracy of the integration.  In addition, the particle time steps are 
quantized in a power-of-two hierarchy of the largest time step.  
The time integration was performed with high enough accuracy to
ensure that the total energy was conserved to better than 0.3\% in
all runs. For all of the models, we followed the time evolution of the density
profile $\rho(r)$, the radial velocity dispersion $\sigma_{r}$, the
tangential velocity dispersion $\sigma_{t}$, and the velocity
anisotropy parameter $\beta$.  All quantities are plotted from the 
softening length, $r = \epsilon$, outward.  We have also explicitly 
checked that our results are not compromized by choices of force softening, 
time-stepping, or opening angle criteria in the treecode.

Results for models A, B, C, D, and E are presented in a 
system of units where
the gravitational constant, $G$, the scale radius of the model, 
$r_{\rm s}$, and the mass within the scale radius, $m(r_{\rm s})$, are all
equal to unity.  With this choice of units, the crossing time at the
scale radius is $t_{\rm cross}(r_{\rm s})=\left[r_{\rm s}^{3} / G m(r_{\rm s})\right]^{1/2}=1$, 
and we adopt it as our time unit.  The half-mass radius of
the models is $r_{\rm h}=(\sqrt{2}+1)r_{\rm s} \simeq 2.41$, and the
orbital timescale at the half-mass radius is 
$t_{\rm h}=2 \pi r_{\rm h} / V_{\rm c}({r_{\rm h}})\simeq 16.6$ time units.
\vspace{1mm}

\begin{figure*}
\centerline{\epsfxsize=4in \epsffile{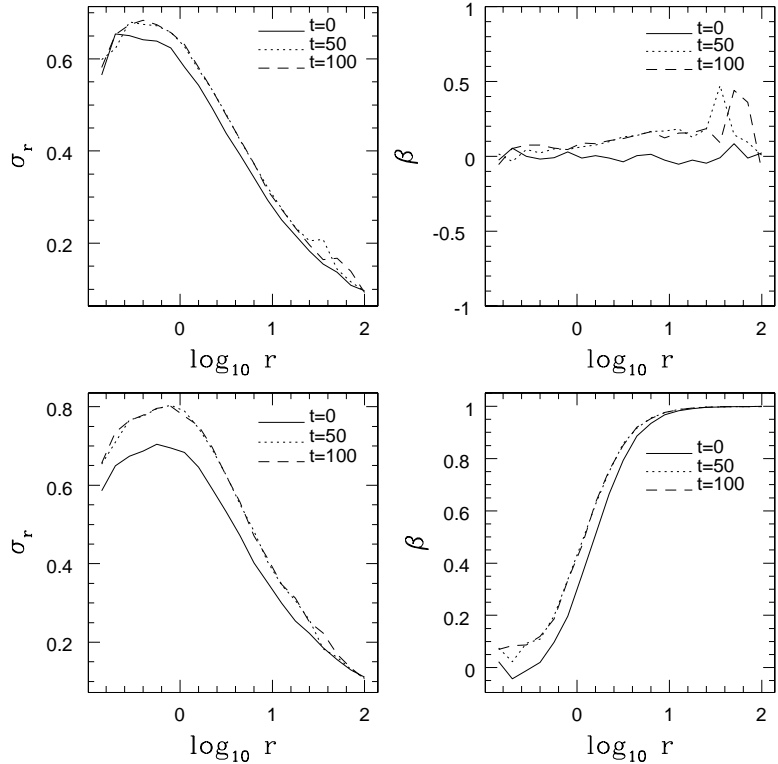}}
\caption{Long-term evolution in the radial velocity dispersion and anisotropy parameter as a
function of time for a Hernquist halo having an isotropic velocity
distribution (model B, {\it top panels}) and a Hernquist halo with
anisotropy radius $r_{\rm a}=(4/3)\,r_{\rm s}$ (model C, {\it bottom panels}).
Both models are simulated with $10^{5}$ particles and were constructed using
the local Maxwellian approximation. 
The solid lines show the initial profiles. The dotted lines show the profiles after 50
crossing times at the scale radius.  The dashed lines show
the profiles after 100 crossing times at the scale radius. The evolution
of the velocity structure is significant over the entire models.
\label{fig2}}
\end{figure*}
\begin{figure*}
\centerline{\epsfxsize=7in \epsffile{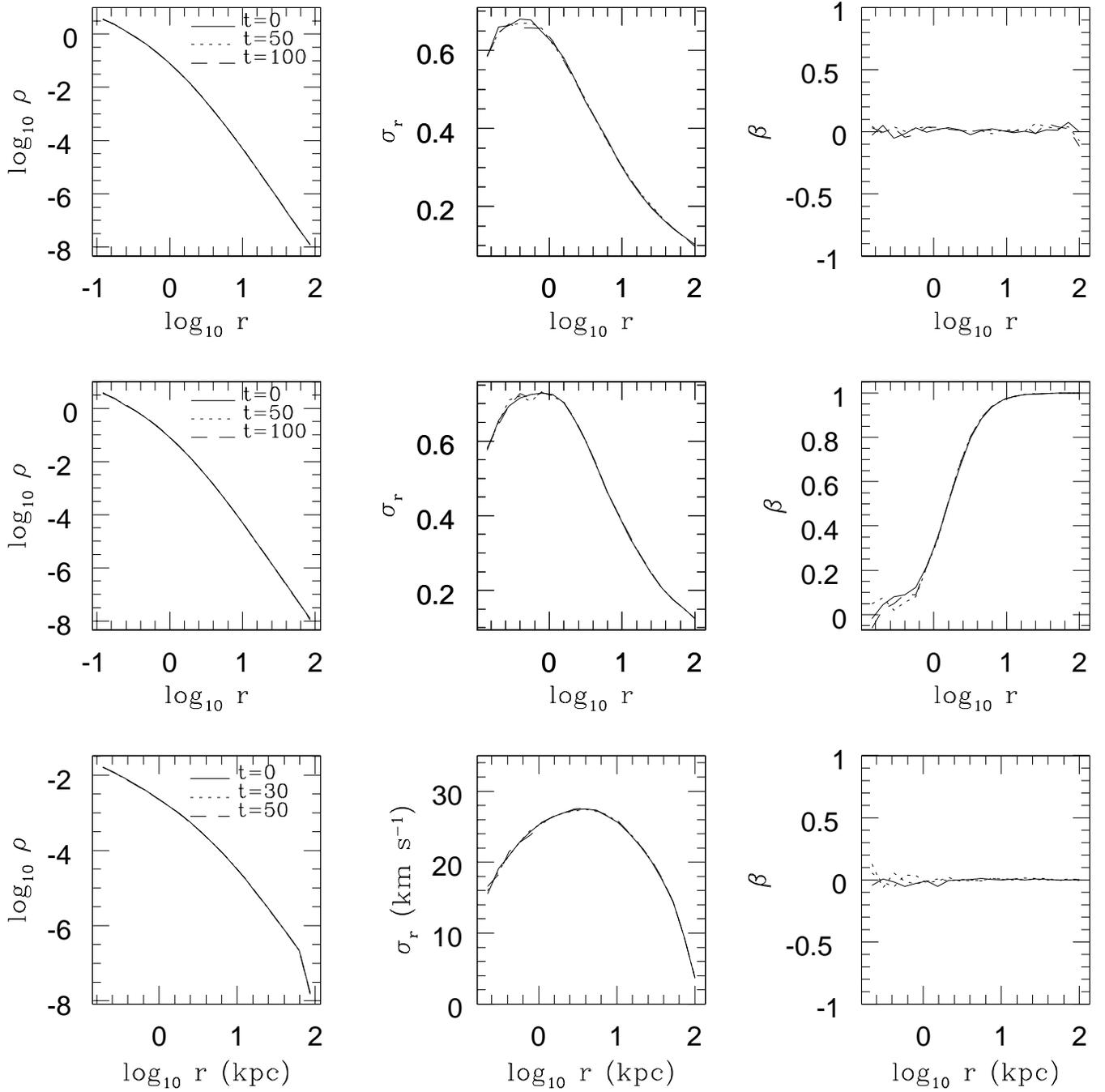}}
\caption{Profile evolution for some of the models generated according to
the procedure described in \S~\ref{sec:initial_conditions}.
The top and middle rows of panels present results
for models D and E, respectively. These differ from models B and C 
(Fig.~\ref{fig2}) only in the initialization procedure used.
The bottom row of panels shows results for model F. 
The elapsed time since the start of the simulations ({\it upper right-hand corners}) 
is given in units of the crossing time at the scale radius.
Virtually no evolution in the density profile or in the velocity
structure can be discerned over the timescales of the simulations.
\label{fig3}}
\end{figure*}
\begin{figure*}
\centerline{\epsfxsize=6in \epsffile{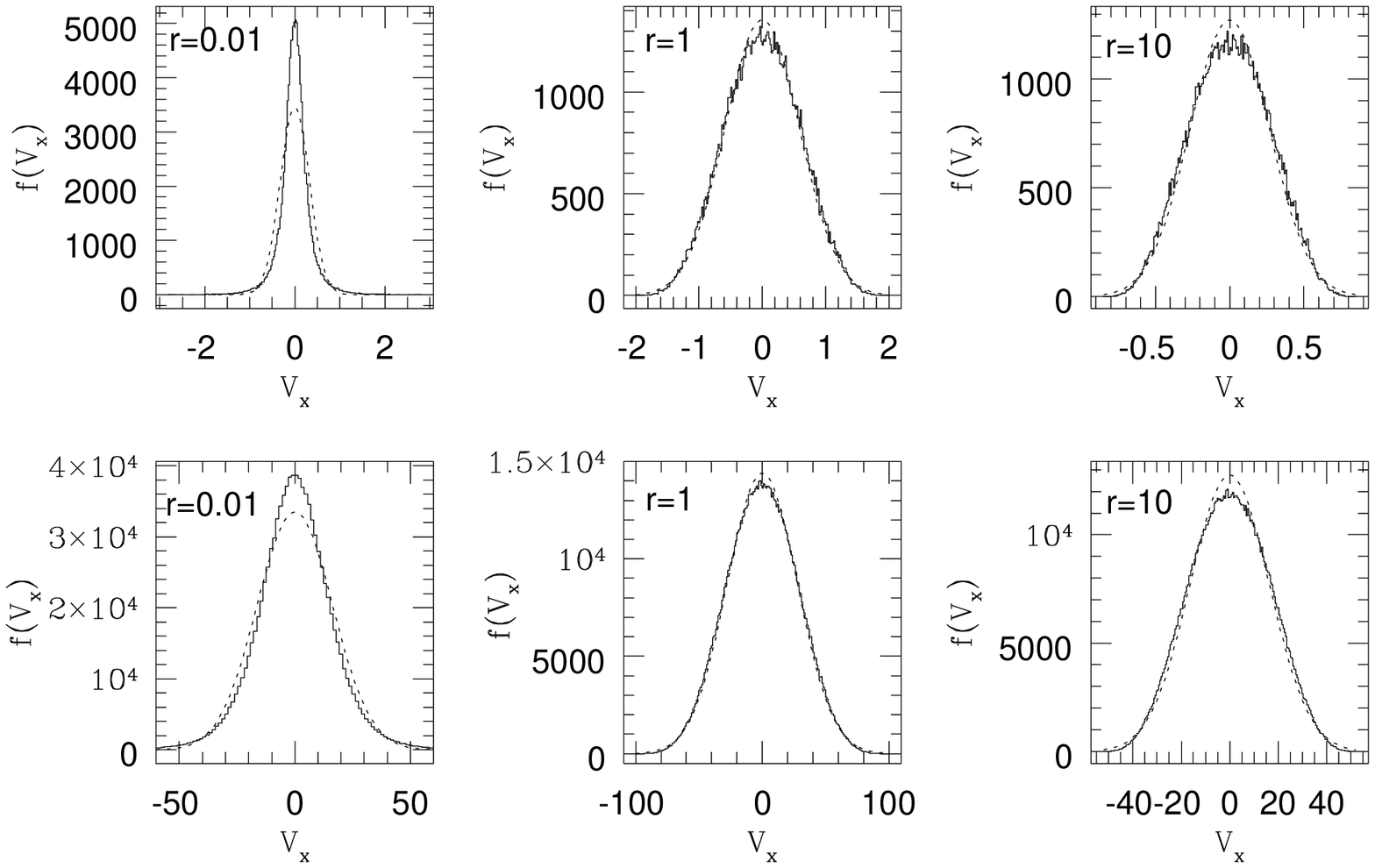}}
\caption{Histograms of the one-dimensional velocity 
distributions for three different distances from the center, 
expressed in units of the scale radius. Results are 
presented for an isotropic Hernquist ({\it top panels}) 
and an isotropic Moore density profile ({\it bottom panels}), 
which correspond to models B and G, respectively.
The dotted lines show the Gaussian velocity distributions with the same 
second moment, as used in the local Maxwellian approximation. 
\label{fig4}}
\end{figure*}

\subsection{Models Generated Using the Local Maxwellian Approximation}

Figure~\ref{fig1} presents results for model A, a high-resolution
realization of an isotropic Hernquist model generated using the local
Maxwellian approximation. The gravitational softening for this model
is $\epsilon=0.015$.  We plot the density profile and the velocity
structure for four different times (initially and after $0.5$, $1$, and $2$
crossing times at the scale radius). The central density cusp almost
immediately (after only $0.5$ crossing times at the scale radius)
becomes very shallow, then starts to fluctuate and relaxes after 
some time to an inner slope much shallower than the initial $\rho\propto r^{-1}$.
The change in the velocity structure is also notable. Both 
the radial and tangential velocity dispersions evolve significantly 
over the course of this run, increasing their initial values by $\sim10\%$.
Note that two-body relaxation or $\sqrt{N}$ mass fluctuations are not
responsible for the observed evolution, since the timescale is too
short: the evolution occurs on a single core crossing time.  The
latter evolution is simply due to the fact that models constructed using
the local Maxwellian approximation are not in equilibrium.

In order to further investigate the effect of the approximate scheme
on the velocity structure of the evolved $N$-body models, we focused
on models B and C. These models were evolved for much longer 
than model A, with the intention of studying the evolution of the velocity structure 
on longer timescales. We ran these models to $t=100$, or approximately $6$ 
orbital times at the half-mass radius, and we used a gravitational softening of
$\epsilon=0.1$.
In Figure~\ref{fig2}, we plot the evolution of the radial velocity dispersion and 
the anisotropy parameter $\beta$. The velocity structure of both models
changes significantly over their entire extent and becomes
radially anisotropic.

\subsection{Models Generated Using the Exact Distribution Function}
 
In order to demonstrate that the evolution seen in models A--C really is a
consequence of the local Maxwellian approximation, we present 
in the top and middle rows of Figure~\ref{fig3} the evolution of models D and E,
respectively. These models have the same number of particles and the same initial density
and velocity profiles as models B and C, respectively, and are
evolved in exactly the same way. The only difference is that the ICs
for models D and E were generated using the procedure in \S~\ref{sec:initial_conditions}.
There is essentially no change in the
density profiles and the velocity structure over the timescales of the
runs.  

The final two simulations demonstrate
the robustness of our adopted procedure for more realistic halo models. 
We ran two simulations of a dwarf galaxy having a virial mass of 
$M_{\rm vir} = 10^{10} \,h^{-1} \Mo$.
In the first simulation, denoted by F, the galaxy followed the NFW density
profile and had a concentration of $c=15$.
In the second simulation, G, the galaxy followed the Moore density profile 
and had a concentration of $c=9.5$.
The crossing time at the scale radius for both models is approximately
$t_{\rm cross}(r_{\rm s}) \sim 0.1$ Gyr, and the orbital timescale at the half-mass radius is 
$t_{\rm h} \sim 3.3$ Gyr for the NFW profile and $t_{\rm h} \sim 3$ Gyr for the 
Moore profile. The scale radius of model F is $ r_{\rm s}=2.94 \,h^{-1}\kpc$, 
and we used a gravitational softening of $\epsilon=0.31 \,h^{-1}\kpc$.
For model G, the scale radius and the chosen softening are
$ r_{\rm s}=4.62 \,h^{-1}\kpc$ and $\epsilon=0.28 \,h^{-1}\kpc$, respectively. 
Finally, we ran our models to $t=50\, t_{\rm cross}(r_{\rm s})$. 
Virtually no evolution in the density profile or in the 
velocity structure can be discerned over the timescales of either simulation, and 
we present the results for the NFW run in the three bottom panels of Figure~\ref{fig3}.
We therefore conclude that $N$-body models generated 
using our algorithm are in equilibrium.

\subsection{Local Velocity Distributions}

How do the exact self-consistent velocity distributions differ from
Maxwellians?  Figure~\ref{fig4} compares the exact one-dimensional 
velocity distribution sampled from the numerically
calculated DF with the Gaussian velocity distribution with the same
second moment.  Results are presented for models B ({\it top panels})
and G ({\it bottom panels}), each for three different radii.
Near the center, the true local velocity distribution in both cases 
is more strongly peaked than a Gaussian, demonstrating that using 
the local Maxwellian approximation is incorrect. The deviation from Gaussianity is
stronger in model B's $\rho(r)\propto r^{-1}$ density cusp than in model G's 
$\rho(r)\propto r^{-1.5}$ cusp: as the inner cusp becomes closer to the $r^{-2}$ profile of a singular
isothermal sphere, the local velocity distribution becomes closer to Gaussian.
As one moves farther out from the center of the system, the
differences between the velocity distributions become smaller, but are
still evident.  At distances close to the scale radius, the two
distributions become closer, and the true velocity distribution starts
to resemble a Gaussian. Beyond the scale radius, 
the trend is that the true velocity distribution is less peaked than a Gaussian.

\section{SURVIVAL OF SUBSTRUCTURE WITHIN COLD DARK 
MATTER HALOS}
\label{sec:evolution}

In this section, we demonstrate the significance of using equilibrium models
for an important application within cosmology:
investigating the evolution and survival of live (mass-losing) 
satellite systems orbiting within a deeper, static host CDM potential.
The NFW density profile is used for both the satellites and the
spherically symmetric static primary potential.  The latter represents a
Milky Way--sized halo with virial mass $M_{\rm prim} = 10^{12} \,h^{-1}
\Mo$ and concentration $c_{\rm prim}=12$.  The mass ratio between the
host and the satellite was chosen to be $M_{\rm prim}/M_{\rm
sat}=1000$.  The satellite was modeled with $N=10^{5}$ particles and a
much higher concentration of $c_{\rm sat}=17$, corresponding to earlier
formation epochs and therefore higher densities of low-mass systems in
CDM models (Eke, Navarro \& Steinmetz 2001). 
Each set of experiments used identical orbits and identical host and 
satellite halos, except that the satellite velocities
were initialized using the two different techniques discussed above.
Here we present results for two sets of experiments 
(see also Hayashi \etal 2003 for similar simulations, but with satellites 
constructed using only the local Maxwellian approximation):

\begin{enumerate}

\item The satellite was placed on an eccentric orbit with apocenter
radius $r_{\rm apo}=7.5\ r_{\rm s,host}$, where $r_{\rm s,host}$ is
the scale radius of the primary halo, and $(r_{\rm apo}/r_{\rm per})=\,$(5:1), 
close to the median ratio of apocentric to
pericentric radii found in high-resolution cosmological $N$-body
simulations (Ghigna \etal 1998).

\item The satellite was placed on a circular orbit with orbital radius
$r_{\rm circ}=3.7\ r_{\rm {s,host}}$. Although circular orbits are a
rarity among cosmological halos, this example allows us to
investigate the effect of the tidal field in a radically different,
non-impulsive regime (the satellite is constantly pruned instead of
undergoing a repeated series of shocks, as in the previous case).
\end{enumerate}

\begin{figure}[t]
\centerline{\epsfxsize=3.5in \epsffile{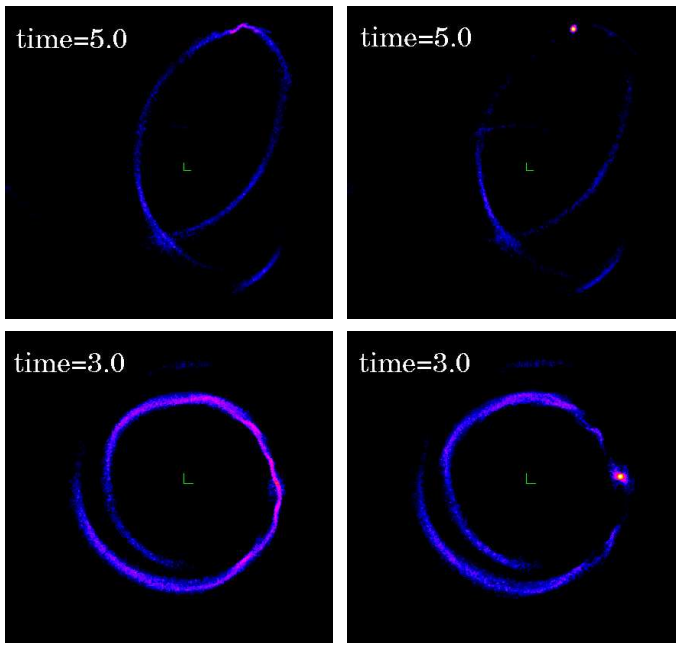}}
\caption{Snapshots of the last stages of the NFW satellites' evolution on 
both orbits inside a Milky Way--sized halo projected onto the orbital plane.  
The satellites constructed using the approximate 
Gaussian scheme are displayed on the left, with the self-consistent satellites
presented on the right.
The color-coded logarithmic density maps are shown (the brighter the color, 
the higher the density), and the elapsed time since the start of the simulation 
({\it upper left-hand corners}) is given in units of the radial orbital period. 
The two top panels show results for the eccentric 5:1 orbit 
with an apocenter radius equal to $r_{\rm apo}=7.5\ r_{\rm s,host}$, 
where $r_{\rm s,host}$ is the scale radius of the host halo, whereas 
the two bottom panels present results for the circular orbit with orbital radius
$r_{\rm {circ}}=3.7\ r_{\rm s,host}$.
After $5$ and $3$ orbital periods, respectively, there is no sign of a self-bound 
core on the left-hand panels, indicative of the fact that the Maxwellian satellites 
have been completely disrupted by the strong tidal field. On the other hand, 
the self-consistent satellites clearly survive complete disruption for the 
same timescales, retaining a self-bound core.
\label{fig5}}
\end{figure}

For both sets of experiments, we followed the time evolution of the
orbit in units of the orbital period, and we present in Figure~\ref{fig5} 
the color-coded logarithmic density maps of the last stages of 
the satellites' evolution. In both pairs of panels, the satellites constructed using the approximate 
Gaussian scheme are displayed on the left, while the self-consistent 
satellites are presented on the right.
The two top panels of Figure~\ref{fig5} show results for 
the eccentric orbit. The satellites lose
mass continuously on account of the strong tidal field, and the
material that has been stripped off forms the familiar tidal tails
that trace the orbital path.  At each pericentric
passage, the satellites experience the strongest tidal shocks which
result in an increase of the mass-loss rate (Gnedin, Hernquist \&
Ostriker 1999; Taffoni \etal 2003). The top left-hand panel shows that 
after approximatelly $5$ orbits, the Gaussian satellite has been fully
disrupted. On the other hand, the evolution of the self-consistent satellite 
was completely different. In this case, we note that the 
satellite was not fully dissolved after $5$ orbits, retaining a 
conspicuous self-bound core ({\it top right-hand panel}).  
Similarly, the two bottom panels of Figure~\ref{fig5} show the 
evolution of the satellites on the circular orbit. The bottom
left-hand panel demonstrates that the satellite generated
with the Maxwellian approximation is completely disrupted after only
three orbits. The self-consistent satellite, on the
other hand, retains a self-bound core that survives the tidal stripping
({\it bottom right-hand panel}). 

\begin{figure}[t]
\centerline{\epsfxsize=3.5in \epsffile{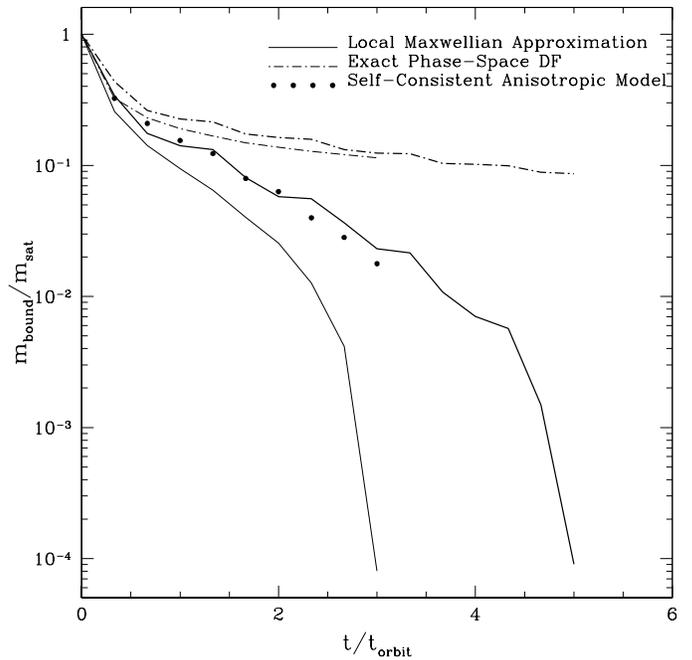}}
\caption{Bound satellite mass as a function of the orbital period 
for the eccentric ({\it thick lines}) and circular ({\it thin lines}) orbits
described in the text.
The solid lines indicate satellite models generated using the local 
Maxwellian approximation, while the dot-dashed lines indicate models 
generated using the exact phase-space DF.  
The large dots correspond to the self-consistent, anisotropic satellite 
placed on the circular orbit.  
The satellites follow the NFW density profile.
Satellites constructed under the assumption that the local velocity field 
is Maxwellian are totally disrupted in a few orbits, whereas the 
self-consistent satellites survive for much longer in the same tidal field.
The anisotropic self-consistent satellite loses mass much more efficiently 
than its self-consistent isotropic counterpart on the same orbit and external tidal 
field. The latter demonstrates that the development of a global bias towards
radial orbits in the Maxwellian satellites is one of the key factors in their 
accelerated disruption.
\label{fig6}}
\end{figure}

In Figure~\ref{fig6} we perform a more quantitative comparison of the 
satellites' evolution, by identifying the mass that remains self-bound 
as a function of time. For that purpose we use the group finder SKID (Stadel 2001),
which has been extensively used to identify gravitationally bound groups
in $N$-body simulations.
The figure shows results for the eccentric ({\it thick lines}) and 
circular ({\it thin lines}) orbits. In both cases, the Maxwellian and 
self-consistent satellites correspond to the solid and 
dot-dashed lines, respectively. The difference between the two 
initialization schemes is apparent in this plot and supports the view 
that the equilibrium model satellites slowly lose mass 
but do not completely disrupt within the timescales of our simulations.

We performed additional numerical experiments with circular orbits of
different radii, as well as eccentric orbits with different
eccentricities, and always found the same result: model satellites
generated under the assumption that their local velocity field is
Maxwellian survive in the tidal field for a much shorter timescale
than their self-consistent counterparts. The former models
are not ``born'' in equilibrium, the rapid expansion of their centers 
makes their density profiles become significantly shallower, and 
their velocity dispersion tensors become more radially biased (Fig. \ref{fig1}).
We find no difference between the mass-loss rate of the Maxwellian satellites
after allowing them to relax for $10$ dynamical times at the
half-mass radius and that experienced by the same satellites
when placed directly in the external tidal field.

In order to demonstrate the importance of a radially anisotropic velocity 
distribution in accelerating the disruption of a satellite,
we have used the procedure of  \S~\ref{sec:initial_conditions} to construct 
a self-consistent anisotropic NFW satellite identical 
to the ones used above and with an anisotropy radius of $r_{\rm a} =(4/3)\, r_{\rm vir}$.
The model was placed on the same circular orbit as before, and we followed  the evolution 
of its bound mass for three orbital periods.
The large dots in Figure~\ref{fig6} present results for this run, indicating that a 
self-consistent, radially anisotropic satellite experiences more efficient
tidal stripping than its isotropic, self-consistent counterpart 
on the same external tidal field and orbit. Particles on more radial orbits spend,
on average, more time at larger radii and are therefore more easily removed by 
the tidal field. On the other hand, models that remain isotropic and retain their 
steep central density cusp throughout their orbital evolution are 
much harder to destroy. A full study of these processes requires a large
suite of numerical simulations covering a wide parameter space, which is clearly beyond
the scope of the present paper.

We refrain from speculating too much about the implications our
results have for the survival of substructure in CDM halos: 
our models ignore the response of the 
host halo to the satellite, which will affect the satellites' susceptibility to 
disruption. We do note, however, that 
the differences we find in the survival times of satellites are of cosmological
relevance. The period of the circular orbit is of the order of $\sim 3$ Gyr.
This means that in one case the satellite gets disrupted in less than $8$ Gyr, 
while in the other case it survives at least for a time
comparable to a Hubble time. In the case of the eccentric orbit, both
satellites would survive for a time comparable to the cosmic age, but
the more pronounced dissolution of the Maxwellian satellite can change dramatically
the morphological evolution of the baryonic component sitting at its
center (Mayer \etal 2001) and, as is also clear from the
top panels of Figure~\ref{fig5}, the
prominence of the tidal streams produced (Helmi \& White 1999;
Johnston, Sigurdsson \& Hernquist 1999; Mayer \etal 2002). 
In addition, while these orbits are more representative of those of
satellites inside a primary halo close to $z=0$, satellites infalling
into the main halo at much higher redshift will have considerably
shorter orbital times (Mayer \etal 2001; Taffoni \etal 2003), and hence
the difference between the two models will clearly be one of survival
versus disruption even for highly eccentric orbits.

\section{SUMMARY}
\label{sec:summary}

In order to understand the physical processes that shape galaxies and
DM halos, it is often more advantageous to study idealized
$N$-body models of isolated galaxies than to try to extract
insight from lower resolution, less controllable cosmological
simulations. Unfortunately, constructing models of isolated galaxies
with specified properties is not straightforward, but in \S~\ref{sec:initial_conditions}
we have described a procedure for generating equilibrium 
$N$-body realizations for a class of spherical galaxy models. 
Despite their simplicity, these models provide very good descriptions 
of the density profiles of both real and simulated galaxies and halos.

A commonly used alternative way of constructing ICs for galaxy models
is by making the local Maxwellian approximation.  This method has the
attraction of being very general and relatively easy to implement.
In recent years it has been used in investigations of the tidal
stripping of satellites, the survival of substructure inside DM
halos, and the effects of bars on halo profiles, among others.
However, it suffers from a dangerous flaw. In \S~\ref{sec:realizations}
we have presented a detailed analysis of the evolution of the   
density profiles and velocity structure of models 
produced using this approximation and have demonstrated that they
are not in equilibrium. Models that are constructed with a 
  central density cusp (e.g., NFW models) and an isotropic velocity dispersion tensor 
relax to a state with a much shallower inner density slope and 
develop a global bias towards radial orbits. 

This spurious evolution can have spectacular consequences, as demonstrated
in  \S~\ref{sec:evolution}, where we investigated the survival times of satellites
orbiting inside a deeper, static host potential. Compared to self-consistent satellites,
Maxwellian satellites experience accelerated mass loss, leading to 
their artificial dissolution in only a few orbits.
This is something that has to be taken into account when the goal
is to explore the evolution of substructure in CDM halos  
(Taffoni \etal 2003; Hayashi \etal 2003). 
Disruption times change significantly with respect to the cosmic age in the two
initialization schemes.
The fact that the self-consistent satellites are very hard to destroy
supports the claim that abundant substructure is a key feature of CDM
models (Moore \etal 1999b; Klypin \etal 1999).  
An extensive parameter survey of the influence of central density slopes, 
concentration parameters, and peak circular velocities on the survival and 
evolution of substructure halos within CDM models will be addressed in a 
forthcoming paper (S. Kazantzidis, L. Mayer, \& B. Moore 2004, in preparation).

\acknowledgements 

We would like to thank Shaun Cole and 
Prasenjit Saha for carefully reading an early version of the paper
and for providing plenty of useful comments and suggestions.
Stimulating discussions with J\"urg Diemand, John Dubinski, Vincent Eke, 
Adrian Jenkins, Lucio Mayer, and Joachim Stadel are greatly acknowledged.
We also wish to thank the anonymous referee for constructive comments on the 
manuscript. S. K. is grateful to the Canadian Institute
for Theoretical Astrophysics for their hospitality during a visit when 
some of this work was completed.
S. K. was funded by a Leverhulme Trust Prize Fellowship award 
held by B. M., and J. M. acknowledges support from the Royal Society. 
The numerical simulations were carried out as part of the Virgo Consortium.

\end{document}